\documentclass[a4paper,11pt]{article}

\usepackage{multicol}
\usepackage{graphicx}
\usepackage{setspace}
\usepackage[
   pdftitle={K X-Ray Energies and Transition Probabilities for  He-, Li and Be-like Praseodymiumin ions}, %%% <- your title
   pdfauthor={J. P. Santos, A. M. Costa, ,  M. C. Martins, P. Indelicato, and F. Parente},         
   pdfkeywords={electrons, atoms, x-ray spectrum},   %%% <- your keywords
   colorlinks,
   linkcolor=blue,
   urlcolor=blue,
   citecolor=blue,
   a4paper=true,
   ]{hyperref}

\newcommand{\abstracttitle}[1]{
 \begin{center}{\Large {\bf #1}}\end{center}
}
\newcommand{\authors}[1]{
 \vspace*{-0.3cm}
 \begin{center} {\bf #1} \end{center}
 \vspace*{-0.3cm}
}
\newcommand{\addresses}[1]{
 \begin{center} {\small #1} \end{center}
}
\newcommand{\synopsis}[1]{
 \begin{center}
 \setstretch{0.75}
 \begin{minipage}[t]{16cm}
   {\footnotesize {\bf Synopsis} #1 }
 \end{minipage}
 \setstretch{1.0}
 \end{center}
}
\newcommand{\abstracttext}[1]{
 \vspace*{-0.3cm}
 \columnsep0.75cm
 \begin{multicols}{2} #1 \end{multicols}
}
\newcommand{\picturelandscape}[2]{
 \vspace*{0.5cm}
 \centerline{
  \includegraphics*[width=7.8cm,angle=#1]{#2}
 }
}
\newcommand{\capt}[2]{
 \vspace*{-0.3cm}
 \begin{center}
 \begin{minipage}[t]{7.8cm} {\small {\bf Figure~#1}.~#2} \end{minipage}
 \end{center}
 \vspace*{0.3cm}
}

\newcommand{\writeto}[1]{
 \hspace*{-2.5mm} \footnote{E-mail: \href{mailto:#1}{#1}}\hspace*{-1.5mm}
}

\begin{document}

\abstracttitle{
K X-Ray Energies and Transition Probabilities for  He-, Li- and Be-like Praseodymium ions
}

\authors{
J. P. Santos$^{\ast}$\writeto{jps@fct.unl.pt},
A. M. Costa$^{\dag}$, 
M. C. Martins$^{\ast}$,
P. Indelicato$^{\S}$\writeto{paul.indelicato@spectro.jussieu.fr},
F. Parente$^{\ast}$\writeto{facp@fct.unl.pt} 
}

\addresses{
$^\ast$ Centro de F\'isica At\'omica, CFA, Departamento de   F\'isica, Faculdade de Ci\^encias e Tecnologia, FCT, Universidade Nova de Lisboa, 2829-516 Caparica, Portugal \\
$^\dag$ Centro de F\'isica At\'omica, CFA, Departamento de   F\'{\i}sica, Faculdade de Ci\^{e}ncias, FCUL, Universidade de   Lisboa, Campo Grande, 1749-016 Lisboa, Portugal \\
$\S$ Laboratoire Kastler Brossel, \'Ecole Normale Sup\'erieure, CNRS, Universit\'e P. et M. Curie -- Paris 6, Case 74; 4, place Jussieu, 75252 Paris CEDEX 05, France}

\synopsis{
Theoretical transition energies and probabilities for  He-, Li and Be-like Praseodymium ions are
  calculated in the framework of the multi-configuration Dirac-Fock
  method (MCDF), including QED corrections. These calculated values are compared to recent
  experimental data obtained in the Livermore SuperEBIT electron beam ion trap facility~\cite{Thorn}. }

\abstracttext{
Highly-charged ions constitute a testbed for the understanding of the electronic
structure of matter, because they can be considered a subclass of atoms in which the basic 
physical parameters take on extreme values.

In the present \textit{ab initio} theoretical work we start from a
Dirac-Fock calculation with Breit interaction included
self-consistently.  Higher-order retardation and one-electron
radiative corrections are also included, and the screening of the
self-energy is evaluated using the Welton approximation.  Correlation
is added within the multiconfiguration Dirac-Fock method (MCDF). In
this framework we have calculated the relativistic transition
energies for the most important He-like, Li-like, and Be-like Praseodymium
K lines, and used them to compute the transition
probabilities. 

Using a preliminary calculation of excitation and ionization  cross sections from
the ions ground configurations, as in our earlier work for silicon ions~\cite{martins2009}, we were able to synthesize a theoretical spectrum, 
which was compared to experimental data obtained in the Livermore SuperEBIT electron beam ion trap facility~\cite{Thorn}.

The more important features in this spectrum are the He-like $1s 2s\,^3S_1 \to 1s^2\,^1S_0$ M1 line,  $1s 2p_{1/2}\,^3P_1 \to 1s^2\,^1S_0$ , and  $1s 2p_{3/2}\,^1P_1 \to 1s^2\,^1S_0$ E1 lines, the Li-like  $1s 2s^2 \,^2S_{1/2} \to 1s^2 2p_{1/2}
\,^2P_{1/2}$ two-electron one-photon line,  and the $1s 2s 2p_{3/2} \,^2P_{3/2} \to 1s^2 2s
\,^2S_{1/2}$ E1 line.

\picturelandscape{0}{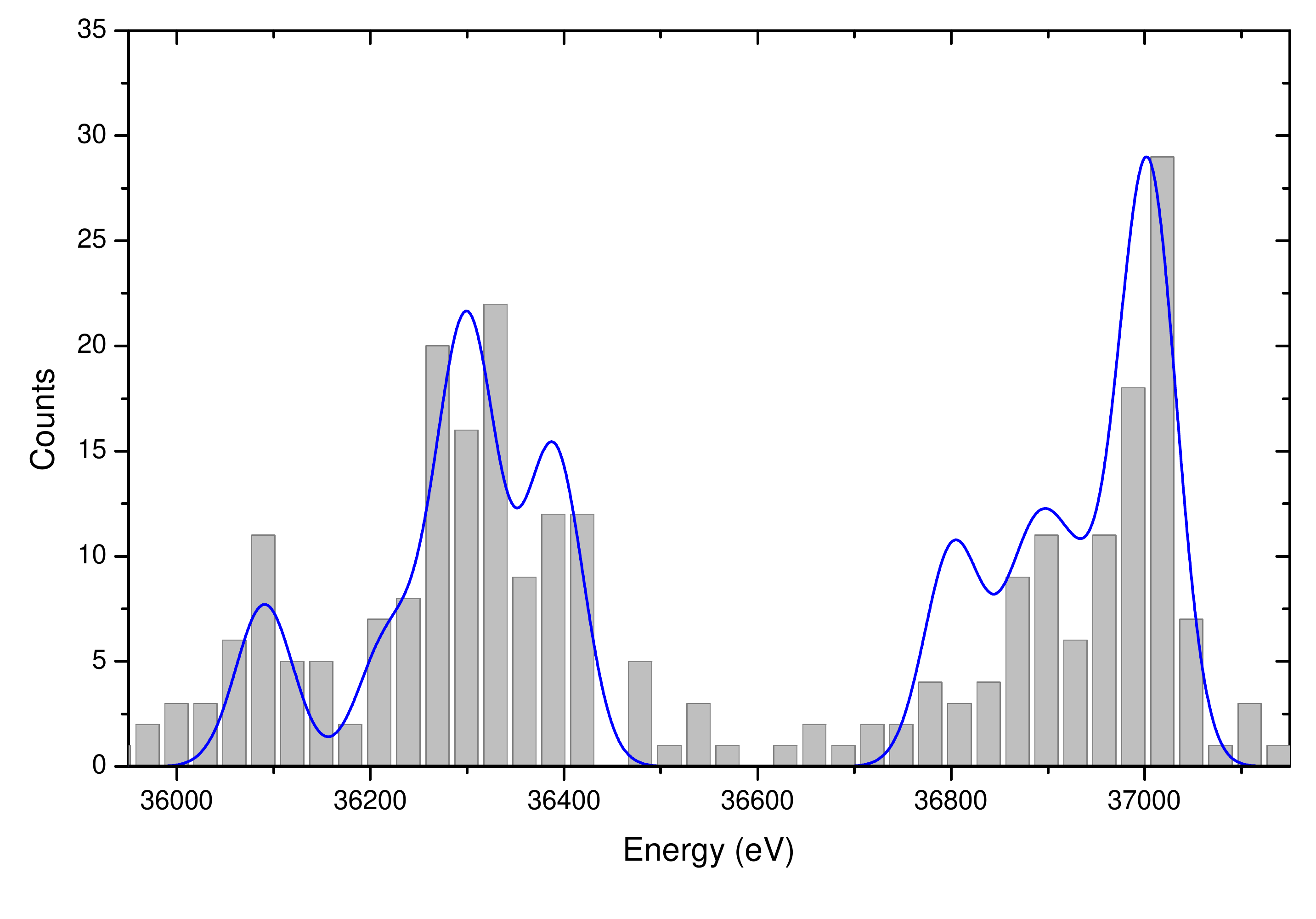} % Use this for pdflatex 
\capt{1}{Synthesized theoretical He-, Li- and Be-like Praseodymium spectrum compared with experimental data from Ref.~\cite{Thorn}.}

For the He-like $1s 2s\,^3S_1 \to 1s^2\,^1S_0$ and $1s 2p_{3/2}\,^1P_1 \to 1s^2\,^1S_0$ line energies we obtained the values 36305.08 keV, and 37003.43 keV, respectively, in very close agreement with the values of the high-precision QED calculations from Ref.~\cite{asyp2005}. Despite the low statistics of the experimental data, our synthesized spectrum is able to account for most of its features. Complete results will be published elsewhere~\cite{santos}.

%========= BIBLIOGRAPHY ============================================ 

}  
% end of the body

\begingroup
\small
\begin{thebibliography}{9}

\bibitem{Thorn} D. B. Thorn {\em et al} 2008 {\em Can. J. Phys.}
{{\bf 86} 241}

\bibitem{martins2009} M. C. Martins {\em et al} 2009 {\em Phys. Rev. A}
{{\bf 80} 032501}

\bibitem{asyp2005} A. N. Artemyev {\em et al} 2005 {\em Phys. Rev. A}
{{\bf 71} 062104}

\bibitem{santos} J. P. Santos {\em et al} 2011 {\bf To be submitted} 
%\href{http://iopscience.iop.org/1742-6596/194/00/001001}{{\bf 194} 001001}

\end{thebibliography}
\endgroup
\end{document}